\documentstyle[12pt]{article}

\newcommand{\beq}{\begin{equation}}
\newcommand{\eeq}{\end{equation}}
\newcommand{\lab}{\protect\label}
\newcommand{\cutoff}{\Lambda_\chi}

\begin{document}

\begin{titlepage}
\begin{flushright}
ROM2F/94/32\\
\end{flushright}
\vspace{0.5cm}
\begin{center}

{
%\Large
\bf
The S parameter in a technicolour model \\
with explicit chiral symmetry breaking}

\vspace*{1.0cm}

\vspace*{1.2cm}
         {\bf E. Pallante\footnotemark
      \footnotetext{ email: pallante@vaxtov.roma2.infn.it\\
      \hspace*{6.5mm}fax: ++39-6-9403-427}} \\
{\footnotesize{ I.N.F.N., Laboratori Nazionali di Frascati,
Via E. Fermi, 00144 Frascati ITALY}} \\
\end{center}

\vspace*{3.0cm}

\begin{abstract}

We derive the S parameter of the electroweak radiative corrections in a
scaled-QCD technicolour and using a Nambu-Jona Lasinio model for
hadronic low energy interactions.  In this framework deviations from low
energy QCD can be quantitatively deduced.  We induce explicit chiral
symmetry breaking in the NJL model through the current-quarks mass term.
It is shown that the prediction for the S parameter can be sensitively
reduced respect to the chiral prediction.

\end{abstract}
\vspace*{1.5cm}
\begin{flushleft}
ROM2F/94/32  \\
hep-ph/9408232
\end{flushleft}
\end{titlepage}
\newpage

The S parameter of the electroweak radiative corrections is defined in
terms of two of the gauge bosons vacuum polarization functions as
\cite{Peskin}

\beq
\alpha S \equiv 4e^2 [\Pi_{33}^\prime (0) - \Pi_{3Q}^\prime (0) ],
\lab{SDEF}
\eeq

where $\alpha$ is the electromagnetic fine structure constant, the index
3 refers to the isospin $SU(2)_L$ current, while the index Q refers to
the $U(1)_{em}$ current.  The primed functions arise from the expansion
in $q^2$ of the $\Pi$ functions defined as

\beq
i\int d^4x~ e^{-iqx}<J^\mu_X(x)J^\nu_Y(0)>\equiv -g^{\mu\nu}
\Pi_{XY}(q^2)+q^\mu q^\nu terms,
\lab{2PF}
\eeq

with XY = 33, 3Q. By neglecting terms of order higher than
 $q^2$ one has:

\begin{eqnarray}
\Pi_{33}(q^2)&\sim& \Pi_{33}(0)+q^2\Pi_{33}^\prime (0),\nonumber\\
\Pi_{3Q}(q^2)&\sim&q^2\Pi_{3Q}^\prime (0),
\end{eqnarray}

where $\Pi_{3Q}(0)=0$ because of the QED Ward identity.

In a technicolour model where isospin and parity are conserved the $S$
parameter can be reexpressed in terms of the isospin-vector and axial
vector two point Green's functions through the relations

\beq
\Pi_{33}={1\over 4}(\Pi_{VV}+\Pi_{AA}), ~~~~\Pi_{3Q}={1\over 2}
\Pi_{VV},
\lab{REL}
\eeq

where $\Pi_{VV}$ and $\Pi_{AA}$ are defined according to eq.
(\ref{2PF}) with XY = VV, AA.

{}From eq. (\ref{SDEF}) and relations (\ref{REL}) the S parameter can be
written in terms of the Vector and Axial two-point functions as:

\beq
S = - 4\pi ( \Pi^\prime_{VV}(0) - \Pi^\prime_{AA}(0) ),
\eeq

where $\Pi_{XY}^\prime (0) = d/dq^2 \Pi_{XY}(q^2)\vert_{q^2 =0}$.

The two-point vector and axial functions can be explicitely derived in
effective quark models {\em \`a la} Nambu-Jona Lasinio (for a review see
also \cite{NJL,MANYNJL}) formulated for hadronic low energy interactions
\cite{ENJL,2point,QR}.  They have been calculated in the ENJL model with
$n_f=2,3$ and in the chiral limit in \cite{2point}, while a calculation
in the non chiral limit with a different from the present approach can
be found in \cite{3point}.  They can be written in the usual VMD form
with scale dependent meson parameters and $Q^2=-q^2$:

\begin{eqnarray}
{1\over -Q^2}\Pi_{VV}(Q^2)&=&- {f_V^2(Q^2) M_V^2(Q^2)\over M_V^2(Q^2)
+Q^2 }
\nonumber\\
{1\over -Q^2}\Pi_{AA}(Q^2) &=&- {f_\pi^2(Q^2)\over Q^2} -
{f_A^2(Q^2) M_A^2(Q^2)\over M_A^2(Q^2) +Q^2 },
\end{eqnarray}

\noindent where in QCD $f_\pi=93.3$ MeV is the pion decay constant,
$f_V$ and $f_A$ are the couplings of the vector meson to the external
vector current and of the axial meson to the external axial current,
$M_V$ and $M_A$ are the masses of the vector and axial mesons.

By substituting $\Pi^\prime_{VV}(0)$ and $\Pi^\prime_{AA}(0)$ in the
definition of the S parameter we obtain:

\beq
S = 4\pi [f_V^2(0) - f_A^2(0)].
\eeq

Using the effective action approach the values at $Q^2=0$ of $f_V^2$
and $f_A^2$ in the chiral limit are \cite{ENJL}:

\begin{eqnarray}
f_V^2(0)&=& {N_c\over 16\pi^2} {2\over 3} \Gamma
\biggl ( 0,{M_Q^2\over \cutoff^2}\biggr )\nonumber\\
f_A^2(0)&=&{N_c\over 16\pi^2} {2\over 3} g_A^2(0)\biggl [
\Gamma \biggl ( 0,{M_Q^2\over \cutoff^2}\biggr ) -
\Gamma \biggl ( 1,{M_Q^2\over \cutoff^2}\biggr ) \biggr ] .
\end{eqnarray}

The expression for S results:

\beq
S = 4\pi \biggl [
{N_c\over 16\pi^2}{2\over 3}\Gamma \biggl ( 0,{M_Q^2\over
\cutoff^2}\biggr )
(1-g_A^2(0))+{N_c\over 16\pi^2} {2\over 3}g_A^2 \Gamma \biggl (
1,{M_Q^2\over \cutoff^2}\biggr )
\biggr ] .
\lab{schiral}
\eeq

The $g_A$ parameter is the mixing parameter between axial and
pseudoscalar mesons given by:

\beq
g_A(0) = {1\over 1+4G_V{M_Q^2\over \cutoff^2}
\Gamma \biggl ( 0,{M_Q^2\over \cutoff^2}\biggr )} ,
\eeq

with $G_V$ the four-quark vector coupling constant
and the $\Gamma$ functions are defined as

\beq
\Gamma (n-2,\epsilon ) = \int_\epsilon^\infty~ dz~ {1\over z}~
e^{-z}~z^{n-2}.
\eeq

The function $\Gamma (0,\epsilon ) = -\ln \epsilon - \gamma_E + {\cal
O}(\epsilon )$ corresponds to the divergent contribution in a
momentum-cutoff regularization scheme, while $\Gamma (1,\epsilon )$
gives the first finite contribution.

The S parameter is a function of the adimensional ratio
$M_Q^2/\cutoff^2$, where $M_Q$ is the infrared cutoff of the effective
techni-meson theory and $\cutoff$ is the ultraviolet cutoff of the
effective techni-meson theory.  By assuming a technicolour model which
is a scaled QCD model (i.e.  by assuming the same adimensional ratio
$M_Q^2/\cutoff^2$ for both QCD and technicolour theories) we will use
the numerical low energy QCD values for $M_Q$ and $\cutoff$ to get a
prediction for the S parameter.

By inserting the numerical values $\cutoff = 1.160$ GeV, $M_Q = 0.265$
GeV and $g_A(0) = 0.61$, with $G_V=1.263$,
 valid in the chiral limit \cite{ENJL} we obtain

\beq
S = N_c \cdot (0.1 + 0.02).
\lab{chiral}
\eeq

The first term, which is the logarithmically divergent term, gives the
bulk of the contribution, while a small correction comes from the second
term which is finite.  The value of eq. (\ref{chiral}) is in good
agreement with the estimation by Peskin and Takeuchi \cite{Peskin}
obtained from the experimental data by using low energy QCD dispersion
relations and rescaling to technicolour energies.

The present parametrization gives the possibility to study a strong
dynamics which is like the low energy QCD dynamics, but which can
include deviations in the mass spectrum.

In what follows we induce explicit chiral symmetry breaking (E$\chi$SB)
in the ENJL model with the addition of the current-quark mass term and
study its effect on the S parameter.

By assuming a technicolour spectrum with $N_d$ isodoublets of
technifermions $(\tilde{u}, \tilde{d})$ two different pictures can be
analysed:

I) $m_{\tilde{u}}=m_{\tilde{d}}\neq 0$, for each isodoublet; the mass
matrix is given by $N_d$ $(2\times 2)$ degenerate subblocks and isospin
is conserved.

II) $m_{\tilde{u}}\neq m_{\tilde{d}}$ in each isodoublet; the mass
matrix is non degenerate and isospin symmetry is broken.

We focus on item $I)$, while item $II)$ is under study.

\section{ S in the non chiral limit}

The NJL effective fermion Lagrangian is

\beq
{\cal{L}} = \bar{q}(\hat{D}-{\cal M})q +
{8\pi^2G_S\over N_c\cutoff^2}
(\bar{q}^aq^b)(\bar{q}^bq^a) -{8\pi^2G_V\over N_c\cutoff^2}
[(\bar{q}_L^a\gamma_\mu q_L^b)(\bar{q}_L^b\gamma_\mu q_L^a)
+L\to R],
\lab{LAGF}
\eeq

where we have introduced the explicit $\chi$SB current-quark mass term;
the non-renormalizable four-fermion scalar and vector interactions are
generated by the integration over the high frequency modes of quarks and
gluons.  $G_S(\cutoff )$ and $G_V(\cutoff )$ are non perturbative
coupling constants and $\cutoff$ is the ultraviolet cutoff of the
effective theory.  Bosonization of the Lagrangian (\ref{LAGF})
introduces scalar, pseudoscalar, vector and axial meson degrees of
freedom: the integration over quarks degrees of freedom gives rise to
the effective meson Lagrangian whose parameters are generated by one
quark-loop calculation.  To derive the S parameter in the presence of
E$\chi$SB mass term one has a priori to take into account two effects:
the solution $M_i$ of the mass-gap equation, which is the pole of the
constituent-quark propagator, becomes a function of the current-quark
mass $m_i$ where $i$ is the flavour index.  After the bosonization
corrections to the vector and axial meson parameters can be expressed in
terms of the masses $M_i$.

\subsection{The Mass-Gap equation}

The mass-gap equation given by the Lagrangian (\ref{LAGF}) defines the
dressed current-quark propagator as the sum of the bare current-quark
propagator with mass $m_i$ and the tadpole contribution generated by the
four-quark interaction with coupling $G_S$. The mass of the dressed
current-quark propagator is then

\beq
M_i = m_i-{1\over 2}\biggl ( {8\pi^2G_S\over N_c\cutoff^2}\biggr )
<\bar{q}_iq_i>
\eeq

where $<\bar{q}_iq_i>$ is given by

\begin{eqnarray}
<\bar{q}_iq_i> &=& -{i}N_c \int~{d^4p\over (2\pi )^4}
{M_i\over p^2-M_i^2}\nonumber\\
&=&-{N_c\over 16\pi^2}~4M_i^3~\Gamma(-1,{M_i^2\over \cutoff^2}).
\end{eqnarray}

The graph of figure 1 shows the solution $M_i(m_i)$ rescaled for
technicolour as a function of the coupling $G_S$ for different values of
the current-quark masses $m_i$, while the graph of figure 2 shows the
solution $M_i(m_i)$ for $G_S=1.216$ which is the value obtained from the
best fit of ref. \cite{ENJL} in the chiral limit. Assuming for the
techni-pion decay constant the value $f_\pi^T =250$ GeV \cite{Peskin},
in a QCD like effective technicolour theory we define the ultraviolet
cutoff $\Lambda_T\simeq 4\pi f_\pi^T\simeq 3$ TeV. The values of both
masses $M_i$ and $m_i$ in the graphs can be thought as QCD values
rescaled by the relation $M_T = M_{QCD}\cdot \Lambda_T/\cutoff$ with
$\cutoff = 1.160$ GeV.

$M_i$ is a sensitively increasing function of the current-quark mass
$m_i$. The perturbative expansion in powers of the current-quark masses
has a limited validity range.

\subsection{The calculation}

To calculate the non chiral corrections to the $f_V$ and $f_A$
parameters which enter $S$ we use the effective action approach, already
used in the chiral limit in ref. \cite{ENJL}.

The bosonization of the Lagrangian (\ref{LAGF}) and the transformation
from current-quarks to constituent-quarks

\beq
Q_L=\xi q_L~~~~~~~~~~~~~~~~~Q_R=\xi^\dagger q_R,
\eeq

with $\xi = \sqrt{U} = \exp (2i/f_\pi~\Phi )$ the square root of the
usual exponential representation of the pseudoscalar meson octet $\Phi$,
leads to the effective low energy action for physical mesons after the
integration over quarks and gluons degrees of freedom:

\beq
e^{i\Gamma_{eff}[H,W^\pm_\mu;v,a,s,p]} = \int {\cal D}G_\mu~
e^{i\int~d^4x~-
{1\over 4}G_{\mu\nu}^2} \int~{\cal D}\bar{Q}{\cal D}Q~
e^{-i\int~d^4x~\bar{Q}D_E Q}.
\lab{EFFAC}
\eeq

$H,W^\pm_\mu$ on the right-end side are the scalar, vector and
axial meson fields which are the {\em auxiliary} fields of the
 bosonized Lagrangian (\ref{LAGF}) and the effective action is
calculated in the presence of external sources $v,a,s,p$.

The fermionic euclidean operator in the constituent-quark base is:

\beq
D_E = \gamma_\mu \nabla_\mu -{1\over 2} (\Sigma -\gamma_5\Delta )
 -H(x).
\lab{OPER}
\eeq

The covariant derivative $\nabla_\mu$

\begin{eqnarray}
\nabla_\mu&=& \partial_\mu +iG_\mu
+\Gamma_\mu -{i\over 2}\gamma_5 (\xi_\mu - W_\mu^-)
-{i\over 2}W_\mu^+
\end{eqnarray}

contains the vector and axial meson fields $W_\mu^+$ and $W_\mu^-$ and
the vector and axial-vector currents

\begin{eqnarray}
\Gamma_\mu&=& {1\over 2}\{\xi^\dagger [\partial_\mu -i(v_\mu +
a_\mu )]
\xi + \xi [\partial_\mu -i(v_\mu - a_\mu )]\xi^\dagger )\}
\nonumber\\
\xi_\mu&=& i\{\xi^\dagger [\partial_\mu -i(v_\mu + a_\mu )]
\xi - \xi [\partial_\mu -i(v_\mu - a_\mu )]\xi^\dagger )\} .
\end{eqnarray}

The scalar fields $\Sigma$ and $\Delta$ are proportional to
the current quark mass matrix ${\cal M}$

\begin{eqnarray}
\Sigma&=&\xi^\dagger {\cal M}\xi^\dagger + \xi{\cal M}^\dagger\xi
\nonumber\\
\Delta&=&\xi^\dagger {\cal M}\xi^\dagger - \xi{\cal M}^\dagger\xi .
\end{eqnarray}

The scalar field $H(x)$ acquires a non-zero vacuum expectation value
(VEV) which is responsible of the spontaneous chiral symmetry breaking
in the chiral limit: $H(x) = <H>+\sigma (x)$
 where the fluctuation $\sigma (x)$ defines the true scalar meson
field. Its VEV is the solution of the equation

\beq
{N_c\cutoff^2\over 8\pi^2G_S}2<H>_i + <\bar{Q}_iQ_i> = 0,
\lab{VEVEQ}
\eeq

which arises from imposing the extremum condition of the effective
action (\ref{EFFAC}) in the absence of external sources and vector
fields, i.e. in the mean field approximation:

\beq
{\delta\Gamma_{eff}\over \delta H}\vert_{H=<H>,\xi=1,W_\mu^\pm =0,
s=p=v=a=0}=0.
\eeq

In the mean field approximation (which implies $\xi =1$) the VEV of the
scalar density of the constituent-quarks $<\bar{Q}_iQ_i>$ coincides with
$<\bar{q}_iq_i>$ of the current-quarks.  Then eq. (\ref{VEVEQ}) implies
that the VEV of the scalar field is related to the masses $M_i(m_i)$
solutions of the mass-gap equation through the scalar density
$<\bar{q}_iq_i>$.  In the presence of current-quark masses $m_i$ the
identity $M_i=m_i+<H>_i$ is implied for each flavour $i$.

The effective low energy action (i.e. for $Q^2<\Lambda_T^2$) for
techni-meson fields is given by the determinant of the fermionic
operator $D_E$.  Because we are interested in the real part of it we can
compute the determinant of the squared operator

\beq
\Gamma_{eff} = -{1\over 2}\ln det (D_E^\dagger D_E)_\epsilon
\eeq

regularized in some scheme. The squared operator $D_E^\dagger D_E$ is a
bosonic differential operator bounded below by the constituent quark
masses squared $\tilde{M}_i^2$:

\beq
D_E^\dagger D_E \equiv -\nabla_\mu\nabla_\mu +\tilde{M}_i^2+E,
\eeq

where, referring to eq. (\ref{OPER}), $\tilde{M}_i{\bf{1}} = 1/2 ~\Sigma
+<H> = {\cal {M}}+<H>$ with $\xi = 1$ in first approximation.  This
shows that the constituent-quark masses $\tilde{M}_i$ coincide with the
current-quark masses solution of the mass gap equation.  Form now on
$\tilde{M}_i\equiv M_i$.

In the loop-expansion approach the effective action is given by the
perturbative series around the free part of the operator $D_E^\dagger
D_E$: $D_E^\dagger D_E \equiv (D_E^\dagger D_E)_0 + \delta =
-\partial_\mu\partial^\mu + M_i^2 + \delta$. This defines the
constituent free quark propagator $-\partial_\mu\partial^\mu + M_i^2$
and the perturbation $\delta$.

In ref. \cite{ENJL} the fermionic determinant has been calculated in the
chiral limit using the heat-kernel expansion in the proper-time
regularization scheme (for details on this technique see \cite{Ball}).
This method has the advantage to preserve gauge covariance at each step
of the expansion.

Consistency with the loop-expansion approach requires that
the infrared cutoff of the heat-kernel expansion of the fermionc
squared operator is the constituent quark mass matrix $M_i{\bf{1}}$,
 both in the chiral and non chiral limit.  In the first case $M_i$
coincides with the VEV of the scalar field:
$M_i{\bf{1}}=<H>_i{\bf{1}}=M_Q{\bf{1}}$, where $M_Q$ appeared in the
calculation (\ref{schiral}) of the S parameter in the chiral limit.

In the non chiral limit a simple result follows: the low energy meson
parameters $f_V$ and $f_A$ are those calculated in the chiral limit with
the consitutent-quark mass $M_Q$ replaced by its non chiral value
$M_i=<H>_i(M_i)+m_i$, where $<H>$ is in turn a function of $M_i$,
solution of the mass-gap equation for each flavour $i$.  The S parameter
of eq. (\ref{schiral}) follows with the same substitution and its
numerical value corresponds to the contribution of one technicolour
isodoublet $(\tilde{u},\tilde{d})$ in the isospin limit.

Figure 3 shows the parameter $S/N_c$ as a function of the
constituent-quark mass $M_i$ in the case of one isodoublet
$(\tilde{u},\tilde{d})$ with degenerate masses.  It is noticeable that
the maximum value of S is for the ratio $M_i/\Lambda_T$, or equivalently
$M_i/\cutoff$, which approximately corresponds to the chiral limit in
low energy QCD, while its numerical value decreases rather sensitively
by increasing the current techni-quark masses.

Non-perturbative values of the current-techni-quark masses can push the
S parameter towards zero.

In the case that all masses of the isodoublets in the model are equal
the formula (\ref{schiral}) is easely generalized to $n_f$ flavours by
multiplying by $n_f/2=n_d$ with $n_d$ the number of isodoublets in the
model.  Although no extra-information is given on the techni-particles
spectrum the present parametrization reproduces in a clear form the
effect of explicitely chiral symmetry breaking.

$q\bar{q}$ states with current masses $m_i>\Lambda_T$ cannot be treated
in this framework.  Alternative Heavy-quark-effective-theory as an
expansion in inverse powers of the mass $m_i$ could be used.  Although,
from the present behaviour one can infer that the higher mass of the
state the lower is its contribution to the S parameter.

If E$\chi$SB terms are relevant, higher dimensional effective fermion
interactions, proportional to the current techni-quark masses and
suppressed by inverse powers of the ultraviolet cutoff, can modify the
leading behaviour of the NJL model. The relevance of higher dimensional
operators proportional to powers of momenta $Q^2$ has been studied in
\cite{QR}.  The presence of higher dimensional terms can extend the
range of validity of the perturbative expansion in powers of the
current-quark masses and further reduce the numerical value of the S
parameter.

\vspace{3.cm}
{\bf{Acknowledgements}}
\vspace{0.8cm}

\noindent I am grateful to Roberto Petronzio for having called my
attention to this problem and for many stimulating and useful
discussions.

\vfill\eject

\centerline {{\bf FIGURE CAPTIONS}}
\vskip2.truecm

\begin{description}

\item[1)] The solution of the mass-gap equation $M_i$ in TeV as a
function of the scalar coupling constant $G_S$ for given values of the
current techni-quark masses $m_i$=0, 0.3, 0.5, 0.8, 1.1 TeV. The values
of $M_i$ and $m_i$ can be thought as QCD values rescaled by the relation
$M_T = M_{QCD}\cdot \Lambda_T/\cutoff$. In particular tha values of the
techni-quark masses $m_i$ in figure correspond to QCD values $m_i$ = 0,
0.1, 0.2, 0.3, 0.4 GeV.

\item[2)] The solution of the mass-gap equation $M_i$ as a function of
the current-quark masses $m_i$ for a given value of $G_S=1.216$ given by
the fit of ref.  \cite{ENJL} in the chiral limit.

\item[3)] The parameter $S/N_c$ as a function of the solution of the
mass-gap equation $M_i$. The maximum is approximately at the value of
$M_i$ obtained in the chiral limit.

\end{description}
\vfill\eject

\end{document}